\pdfoutput=1
\documentclass[fleqn,10pt]{wlscirep}

\usepackage{multicol}
\usepackage[version=3]{mhchem}
\usepackage{lipsum,graphicx,float}
\usepackage{xcolor, colortbl}
\newcounter{firstbib}
\usepackage{epsfig}
\usepackage{sidecap}
\usepackage{subfigure}
\usepackage{tabularx}
\usepackage[normalem]{ulem}
\usepackage{hyperref}

\makeatletter
\newcommand\footnoteref[1]{\protected@xdef\@thefnmark{\ref{#1}}\@footnotemark}
\makeatother
\begin{document}

\title{A unique hot Jupiter spectral sequence with evidence for compositional diversity}

\author[1,*]{Megan Mansfield}
\author[2]{Michael R. Line}
\author[3]{Jacob L. Bean}
\author[4]{Jonathan J. Fortney}
\author[5]{Vivien Parmentier}
\author[2]{Lindsey Wiser}
\author[6]{Eliza M.-R. Kempton}
\author[7]{Ehsan Gharib-Nezhad}
\author[8]{David K. Sing}
\author[9]{Mercedes L{\'o}pez-Morales}
\author[10]{Claire Baxter}
\author[10]{Jean-Michel D{\'e}sert}
\author[11]{Mark R. Swain}
\author[11]{Gael M. Roudier}

\affil[1]{Department of Geophysical Sciences, University of Chicago, 5734 S. Ellis Avenue, Chicago, IL 60637, USA}
\affil[2]{School of Earth and Space Exploration, Arizona State University, Tempe, AZ 85281, USA}
\affil[3]{Department of Astronomy \& Astrophysics, University of Chicago, Chicago, IL 60637, USA}
\affil[4]{Department of Astronomy and Astrophysics, University of California, Santa Cruz, CA 95064, USA}
\affil[5]{Department of Physics, University of Oxford, Oxford, OX1 3PU, UK}
\affil[6]{Department of Astronomy, University of Maryland, College Park, MD 20742, USA}
\affil[7]{NASA Ames Research Center, Moffett Field, CA 94035, USA}
\affil[8]{Department of Physics \& Astronomy, Johns Hopkins University, Baltimore, MD 21218, USA}
\affil[9]{Center for Astrophysics \textbar~Harvard \& Smithsonian, Cambridge, MA 01238, USA}
\affil[10]{Anton Pannekoek Institute for Astronomy, University of Amsterdam, 1098 XH Amsterdam, The Netherlands}
\affil[11]{NASA Jet Propulsion Laboratory, California Institute of Technology, Pasadena, CA 91109, USA}
\affil[*]{Corresponding author:  \href{mailto:meganmansfield@uchicago.edu}{meganmansfield@uchicago.edu}}

%\keywords{keywords}

%CURRENT TOTALS
%Abstract: 250 words
%Total: ~2050 words
%35 references

\begin{abstract}

The emergent spectra of close-in, giant exoplanets ("hot Jupiters") are expected to be distinct from those of self-luminous objects with similar effective temperatures because hot Jupiters are primarily heated from above by their host stars rather than internally from the release of energy from their formation \cite{Showman2020}. Theoretical models predict a continuum of dayside spectra for hot Jupiters as a function of irradiation level, with the coolest planets having absorption features in their spectra, intermediate-temperature planets having emission features due to thermal inversions, and the hottest planets having blackbody-like spectra due to molecular dissociation and continuum opacity from the $\mathrm{H}^-$ ion \cite{Fortney2008,Parmentier2018,Arcangeli2018}. Absorption and emission features have been detected in the spectra of a number of individual hot Jupiters \cite{Kreidberg2014,MikalEvans2020}, and population-level trends have been observed in photometric measurements \cite{Beatty2014,Triaud2014,Kammer2015,Zhou2015,Keating2019,Beatty2019,Baxter2020,Garhart2020,Dransfield2020}. However, there has been no unified, population-level study of the thermal emission spectra of hot Jupiters such as has been done for cooler brown dwarfs \cite{Manjavacas2019} and transmission spectra of hot Jupiters \cite{Sing2016}. Here we show that hot Jupiter secondary eclipse spectra centered around a water absorption band at 1.4~$\mu$m follow a common trend in water feature strength with temperature. The observed trend is broadly consistent with model predictions for how the thermal structures of solar-composition planets vary with irradiation level, but inconsistent with the predictions of self-consistent one-dimensional models for internally-heated objects. This is particularly the case because models of internally-heated objects show absorption features at temperatures above 2000~K, while the observed hot Jupiters show emission features and featureless spectra. Nevertheless, the ensemble of planets exhibits some degree of scatter around the mean trend for solar composition planets. The spread can be accounted for if the planets have modest variations in metallicity and/or elemental abundance ratios, which is expected from planet formation models \cite{Mordasini2016,AliDib2017,Madhu2017,Cridland2019}.
\end{abstract}

\flushbottom
\maketitle

\thispagestyle{empty}
\begin{multicols}{2}

We performed a statistical analysis of 19 hot Jupiter secondary eclipse spectra obtained with the Wide Field Camera 3 (WFC3) instrument on the \textit{Hubble Space Telescope} (\textit{HST}) using the G141 grism between $1.1$ and $1.7$~$\mu$m. This bandpass is primarily sensitive to water vapor in exoplanet atmospheres, and the largest molecular feature in this wavelength range is a water vapor absorption band centered at about 1.4~$\mu$m. Over the last decade a large sample of exoplanets have been observed using WFC3+G141 to understand their atmospheric water abundances \cite{Sing2016, Tsiaras2018}, and it has become an important tool in understanding exoplanet atmospheres. 

\begin{figure*}[htbp]
    \centering
    \includegraphics[width=\linewidth]{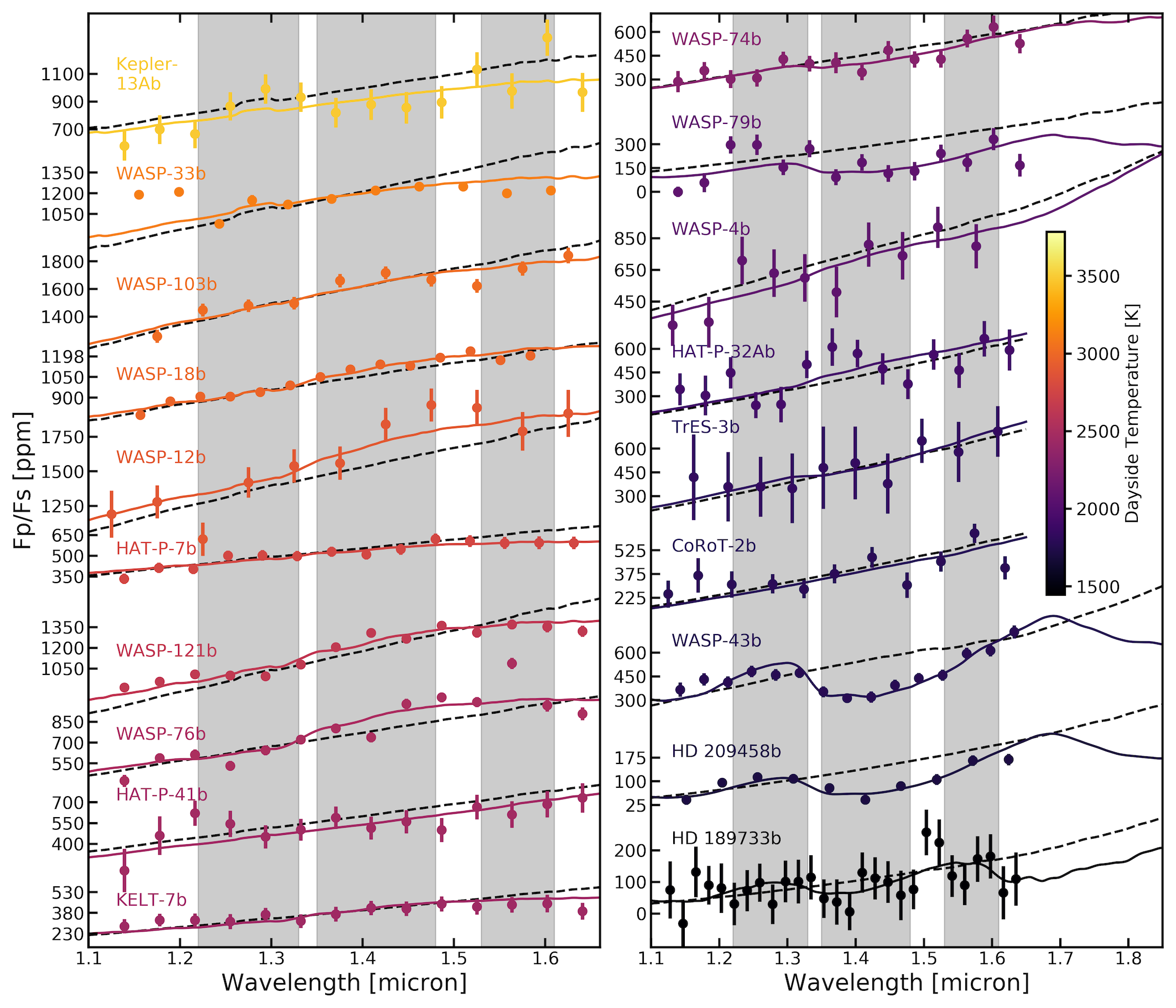}
    \caption{Secondary eclipse spectra showing the planet-to-star flux ratio ($F_{p}$/$F_{s}$) as a function of wavelength for all 19 hot Jupiters considered in this study. Data sets are colored by dayside temperature, which is measured as described in the Methods and shown by the colorbar. Solid lines indicate interpolations from our solar composition fiducial model grid (see the Methods section for a description), while dashed lines indicate best-fit blackbodies. Shaded regions indicate the ``out-of-band'' and ``in-band'' regions used to calculate the water feature strength ($S_{H_{2}O}$) for each observed secondary eclipse spectrum. Note that for several data sets, the error bars are smaller than the point size.}
    \label{fig:wowlookatthedata}
\end{figure*}

We analyzed six new data sets following the data reduction procedure outlined in the Methods. We also performed a reanalysis of the spectrum of one planet, Kepler-13Ab. We combined these seven new analyses with twelve results from the literature to form a complete sample of planets observed with \textit{HST}/WFC3+G141 in this wavelength region. Supplementary Table~1 contains detailed information on each of the twelve literature results we considered. The planets in this study have observed dayside temperatures in the \textit{HST}/WFC3+G141 bandpass between $1450-3500$~K and radii between $0.9-2.0$ Jupiter radii. The full set of 19 spectra are shown in Figure~\ref{fig:wowlookatthedata}.

Manjavacas et al. (2019)\cite{Manjavacas2019} presented the spectra of 10 of the hot Jupiters in this study  but did not examine in detail the physical causes for the observed spectral features. Melville et al. (2020)\cite{Melville2020} similarly examined the spectra of a different subset of 12 hot Jupiters but only analyzed them in the context of models with fixed temperature-pressure (T-P) profiles, with no feedback between the T-P profile and the chemistry. Here we expand on these studies by doubling the sample of hot Jupiter secondary eclipse spectra and comparing the spectra to a grid of models with fully consistent T-P profiles to understand in detail what drives their feature strengths. Because our models combine a set of basic self-consistent assumptions which are expected to hold true for hot Jupiters (e.g., energy balance in the atmosphere and thermochemical equilibrium\cite{Lothringer2018}) with a complete set of relevant opacities, we can use them to create self-consistent predictions for hot Jupiter spectra, which can then be compared to the observed data.

Baxter et al. (2020)\cite{Baxter2020} presented an analysis of changes in \textit{Spitzer Space Telescope} emission photometry with temperature and also examined a subset of planets observed with \textit{HST}. However, because this study focused on broadband photometry, they were only able to give broad constraints on the hot Jupiter population, such as that high C/O ratios are disfavored. This study expands on that work by uniformly analyzing all \textit{HST} thermal emission spectra and performing a more comprehensive analysis of their compositional diversity.

We created a grid of cloud-free irradiated 1D self-consistent radiative-convective-thermochemical equilibrium models to compare to the dayside HST/WFC3+G141 thermal emission observations. These models were created using the Sc-CHIMERA framework \cite{Arcangeli2018,Kreidberg2018,Mansfield2018,Piskorz2018,Zalesky2019,Baxter2020} which includes a broad array of opacity sources that are important for the temperature range explored here, including atomic and ionic opacities that are relevant at the high temperatures of ultra-hot Jupiters \cite{Lothringer2018}. A full description of the models and complete list of opacities can be found in the Methods.

\begin{figure}[H]
    \centering
    \includegraphics[width=0.9\linewidth]{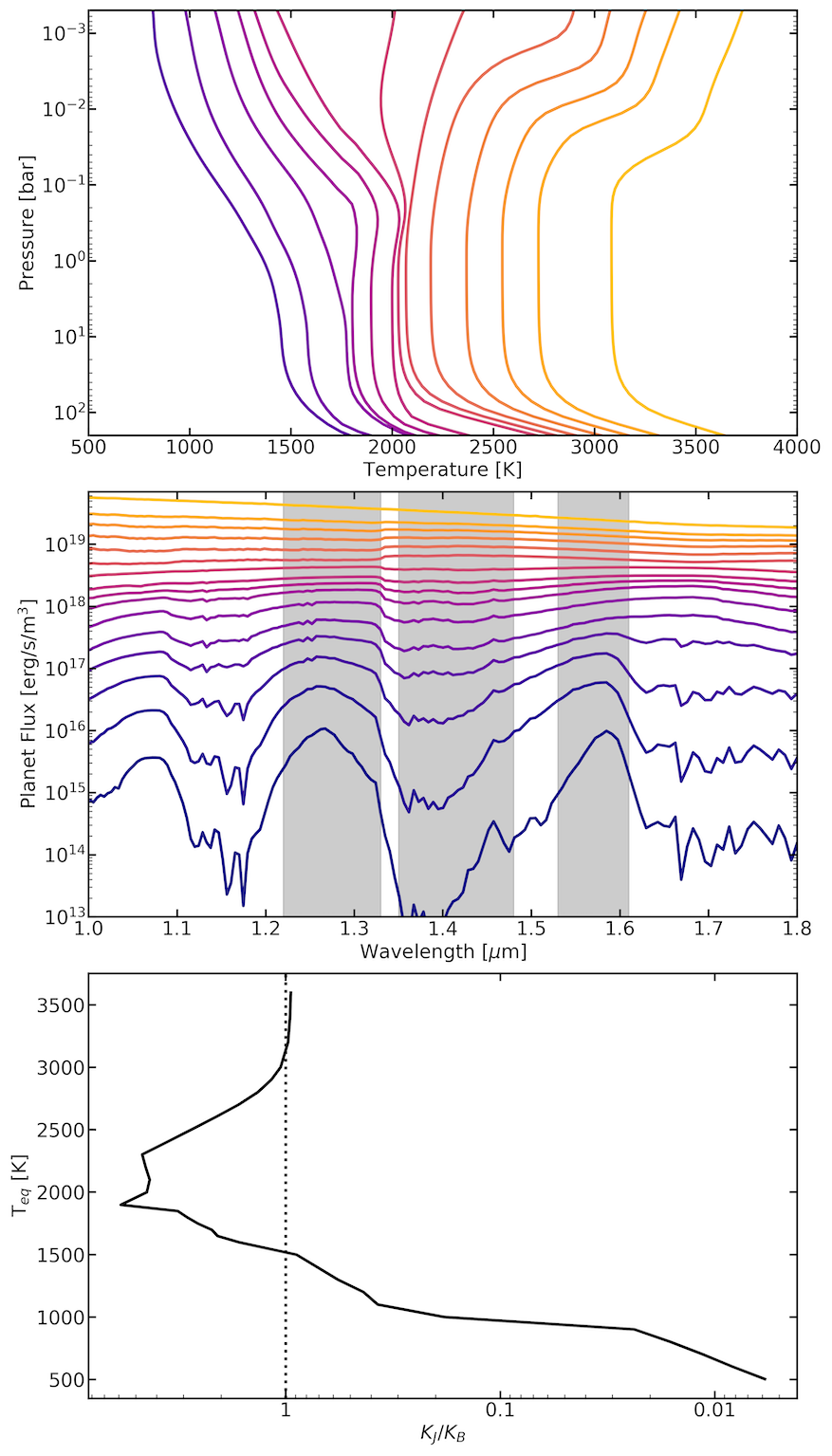}
    \vspace{-2mm}
    \caption{\textbf{(a)} Temperature-pressure (T-P) profiles and \textbf{(b)} resulting dayside planet fluxes for the fiducial model grid, which covers approximately the same range of temperatures as spanned by the observations. The full model specifications are detailed in the Methods. The fiducial model uses a 5300~K stellar effective temperature, a solar composition planetary atmosphere ($\left[\frac{\mathrm{M}}{\mathrm{H}}\right]=0.0$ and $\frac{\mathrm{C}}{\mathrm{O}}=0.55$), a planetary gravity of 10~m/s$^{2}$, and a planet internal temperature of 150~K. Blue and yellow lines show models with the coolest and warmest irradiation temperatures, respectively. The model grid spans a range of irradiation temperatures between $500-3600$~K, with step sizes of $50-200$~K. For clarity, only every other model in the grid is shown here. Grey shaded bands indicate the ``out-of-band'' and ``in-band'' regions used to calculate the water feature strength ($S_{H_{2}O}$) for each model. \textbf{(c)} Ratio of the absorption mean opacity ($\kappa_{J}$) to the Planck mean opacity ($\kappa_{B}$) as a function of equilibrium temperature ($T_{eq}$) in these models at a pressure of $10^{-2}$~bar, assuming zero albedo and full redistribution. This ratio describes the relative efficiency of heating vs. cooling in the models \cite{Lothringer2019}, and a ratio of $\kappa_{J}$/$\kappa_{B}>1$ generally indicates the presence of a thermal inversion in the T-P profile. This panel is plotted with temperature on the y-axis for ease of comparison to Figure~\ref{fig:colormag}.}
    \label{fig:models}
\end{figure}

Figure~\ref{fig:models} shows the T-P profiles and resultant secondary eclipse spectra for our fiducial model, which uses system parameters for a standard hot Jupiter (stellar effective temperature $T_{\mathrm{eff}}=5300$~K, planetary gravity $g=10$~m/s$^{2}$, planetary metallicity $\left[\frac{\mathrm{M}}{\mathrm{H}}\right]=0.0$, planetary carbon-to-oxygen abundance ratio $\frac{\mathrm{C}}{\mathrm{O}}=0.55$, and planetary internal temperature $T_{\mathrm{int}}=150$~K). Models at different temperatures were created by scaling the incident stellar flux to match the specified irradiation temperature. Figure~\ref{fig:models} also shows the ratio of the absorption mean opacity ($\kappa_{J}$) to the Planck mean opacity ($\kappa_{B}$) as a function of equilibrium temperature at a pressure of $10^{-2}$~bar, which is approximately the photospheric pressure in the \textit{HST}/WFC3 bandpass (see the Methods for a full description of these opacities). This ratio describes the relative efficiency of stellar absorption vs. thermal re-radiation at that layer in the planet's atmosphere \cite{Lothringer2019}.

In addition to the fiducial model grid, we created models with a variety of atmospheric/system parameters to see how individual parameters impact the 1D vertical structure and resulting population level trends observed in the dayside emission spectra. We examined models with stellar $T_{\mathrm{eff}}=3300$~K, 4300~K, 6300~K, 7200~K, and 8200~K; planetary gravity, $g=1$~m/s$^{2}$ and 100~m/s$^{2}$; metallicity, $\left[\frac{\mathrm{M}}{\mathrm{H}}\right]=-1.5$ and 1.5; and $\frac{\mathrm{C}}{\mathrm{O}}=0.01$ and 0.85. We also included a model where the internal temperature varies with the planetary irradiation temperature to capture the internal entropy change that could be the cause of the hot Jupiter radii inflation \cite{Thorngren2019}. Furthermore, we tested models in which the TiO and VO opacity were removed {\it ad hoc} until temperatures above 2000~K, 2500~K, or 3000~K in order to simulate cold-trapping in cooler regions of the atmosphere \cite{Parmentier2013,Parmentier2018,Beatty2017} (see the Methods section for a full description of these models). For all of these models, only one parameter was varied at a time while the other parameters were held fixed to the values in the fiducial model (e.g., a slice along a given parameter dimension).

Our models predict three primary spectral regimes. At the lowest dayside temperatures ($T_{\mathrm{day}}<2100$~K for the fiducial model), the models exhibit absorption features due to monotonically decreasing temperature profiles. At intermediate temperatures ($2100$~K~$<T_{\mathrm{day}}<3000$~K for the fiducial model), the modeled thermal structures exhibit a rising temperature with increasing altitude (decreasing pressure) due to the gas-phase onset of TiO and VO which push $K_{J}$/$K_{B}>1$, in turn causing emission features. At the highest temperatures ($T_{\mathrm{day}}>3000$~K for the fiducial model), the models still show strong thermal inversions (becoming stronger primarily due to the dissociation of water, an efficient coolant) but the resulting secondary eclipse spectra are relatively featureless because of a combination of high-temperature effects such as molecular dissociation and the onset of H$^{-}$ opacity, which cause all the WFC3+G141 wavelengths to probe the same altitude/pressure level, hence brightness temperature  \cite{Parmentier2018,Lothringer2018,Arcangeli2018,Kreidberg2018,Mansfield2018}. The exact temperatures of the transitions between these regimes, as well as the strength of absorption and emission features present in the models, depend on the parameters of each set of models.

For both the models and the population of 19 observed hot Jupiters, we examined the degree of absorption or emission observed in the water feature at 1.4~$\mu$m, the primary feature in the \textit{HST}/WFC3+G141 bandpass. We quantified their deviation from a blackbody using an \textit{HST} water feature strength metric, which is illustrated in Supplementary Figure~\ref{fig:wowlookatthedata}. For each data set, we first fit a blackbody to the two ``out-of-band'' regions of the spectrum, which have wavelengths of $1.22-1.33$~$\mu$m and $1.53-1.61$~$\mu$m and are defined based on where the models in Figure~\ref{fig:models} show minimal water opacity. The temperature of this blackbody is referred to throughout this paper as the observed dayside temperature ($T_{\mathrm{day}}$) in this bandpass. The water feature strength is then defined as
\begin{equation}
    S_{H_{2}O}=\log_{10}\left(\frac{F_{B,in}}{F_{obs,in}}\right), 
    \label{eq:strength}
\end{equation}
where $F_{B,in}$ and $F_{obs,in}$ are the flux of the fitted blackbody and the observed data, respectively, in the ``in-band'' region shown in Supplementary Figure~\ref{fig:colorcalc}. The ``in-band'' wavelength region extends from $1.35-1.48$~$\mu$m and captures the center of the primary water band observed in the \textit{HST}/WFC3 bandpass. The shaded regions in Figure~\ref{fig:wowlookatthedata} show the extent of the ``out-of-band'' and ``in-band'' regions. From this definition, $S_{H_{2}O}$ will have a positive value when a water feature is observed in absorption, a negative value when a feature is observed in emission, and a value of zero if a blackbody is observed. We note that we use $S_{H_{2}O}$ here instead of the traditional infrared J- and H-bands because the J- and H-bands exclude the strongest part of the water band at $\approx1.4$~$\mu$m. Therefore, the $S_{H_{2}O}$ index we define gives us greater sensitivity to weak water features that may only produce significant deviation from a blackbody at the center of this absorption band.

\begin{figure*}
    \centering
    \includegraphics[width=\linewidth]{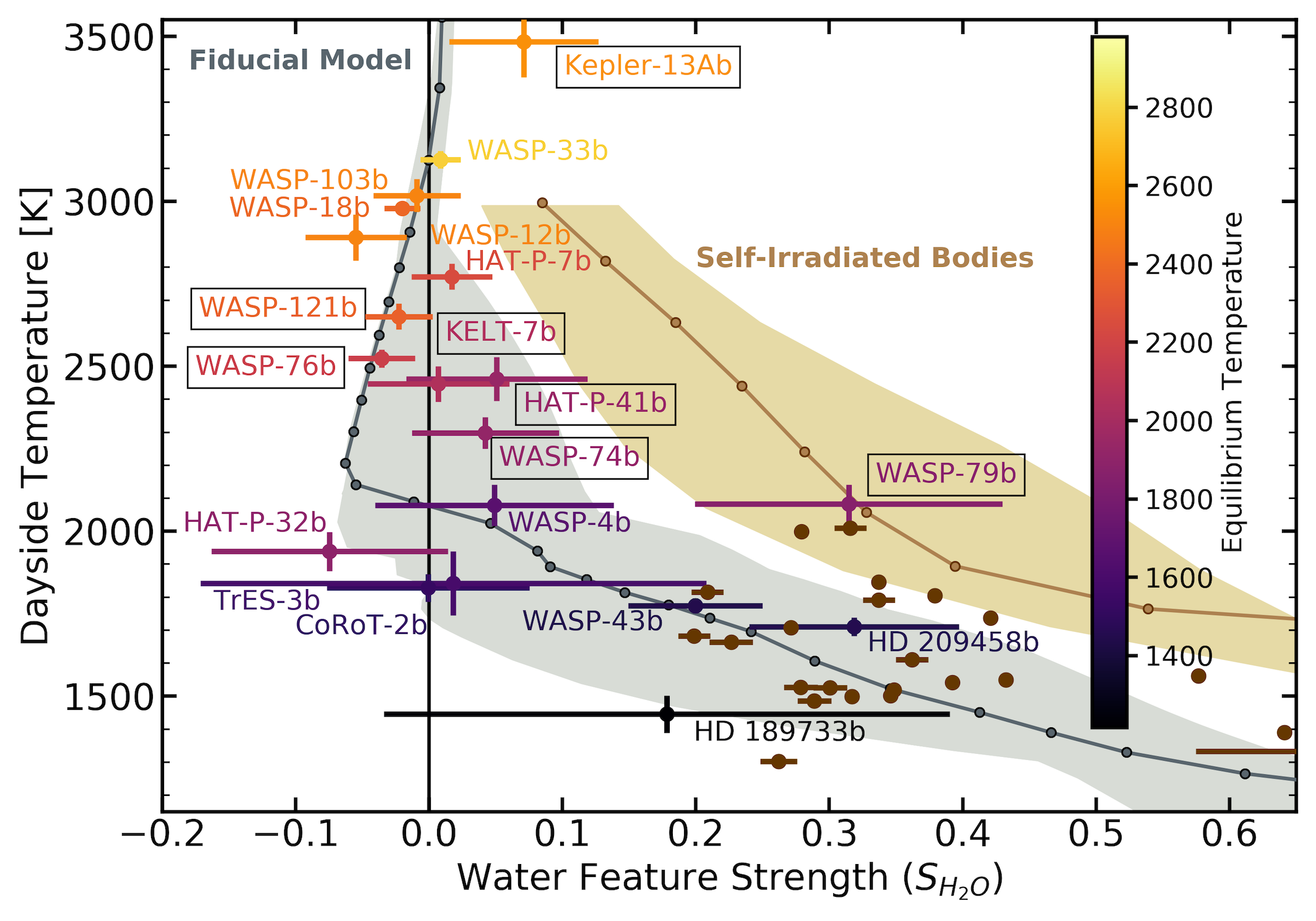}
    \caption{\textit{HST} water feature strength diagram comparing observed secondary eclipse spectra to the model predictions in Figure~\ref{fig:models}. The y-axis shows the temperature of a blackbody fit to the ``out-of-band'' regions defined in Supplementary Figure~\ref{fig:colorcalc}, which is the observed dayside temperature $T_{\mathrm{day}}$. The x-axis shows the strength of the observed feature in the water band at 1.4~$\mu$m compared to this blackbody, as defined by Equation (1). Featureless, blackbody-like spectra have $S_{H_{2}O}=0$ and absorption/emission features have positive/negative values of $S_{H_{2}O}$, respectively. The gray line and points show the fiducial hot Jupiter models pictured in Figure~\ref{fig:models}. The light gray shaded region shows the full range of hot Jupiter model predictions assuming different values for the stellar effective temperature; the temperature where TiO opacity becomes important; and the planet gravity, C/O ratio, metallicity, and internal heat. Similarly, the brown line and points show the fiducial self-luminous object models, and the tan shaded region shows the full range of self-luminous models assuming different values for the planet gravity and metallicity. Colored points with $1\sigma$ error bars show all planets with \textit{HST}/WFC3 spectra, and boxes around planet names indicate new data reductions in this publication. The color scale indicates the planetary equilibrium temperature. The error bars include uncertainties in the stellar effective temperature. Brown points show isolated brown dwarf spectra observed with \textit{HST}/WFC3 from Manjavacas et al. (2019)\cite{Manjavacas2019}.}
    \label{fig:colormag}
\end{figure*}

Figures~3 and 4 show the observed \textit{HST} water feature strengths for the sample of 19 hot Jupiter emission spectra compared to those of the models. Supplementary Table~2 lists the value of $S_{H_{2}O}$ for each planet. Figure~\ref{fig:colormag} shows that the observed \textit{HST}/WFC3 feature strengths generally fall within the region of parameter space spanned by the models, with almost all of the planets fully within the predicted spread of the models. The models considered here assume elemental abundance ratios that fall within the range of commonly expected outcomes from planet formation models \cite{Mordasini2016,AliDib2017,Madhu2017,Cridland2019}. We find that varying parameters in these simple models can explain the observed hot Jupiter population without having to appeal to less likely outcomes of planet formation (e.g., C/O$>1$ \cite{Mordasini2016,AliDib2017,Cridland2019}) or exotic chemistry. 

In order to demonstrate the difference between models of hot Jupiter atmospheres (which are primarily irradiated from above by their host stars) and self-luminous atmospheres (which are primarily heated from below by the object's interior), we created a separate model grid of self-luminous, cloud-free models using the same Sc-CHIMERA code setup. These models used an identical set of parameters to the hot Jupiter models, with the exception of irradiation from within the body instead of from an exterior star. The fiducial self-luminous models had $g=1000$~m/s$^{2}$, $\left[\frac{\mathrm{M}}{\mathrm{H}}\right]=0.0$, and $\frac{\mathrm{C}}{\mathrm{O}}=0.55$. We also created models with metallicities of $\left[\frac{\mathrm{M}}{\mathrm{H}}\right]=-1.0$ and $1.0$. Additionally, while a gravity of $g=1000$~m/s$^{2}$ is typical for a brown dwarf, we created grids with $g=100$~m/s$^{2}$ and $g=10$~m/s$^{2}$ for direct comparison to lower-gravity hot Jupiters.

We used Equation (1) to derive water feature strengths for the grid of self-luminous models. Figure~\ref{fig:colormag} shows these self-luminous water feature strengths compared to those from the hot Jupiter models and hot Jupiter observations, as well as water feature strengths derived from the \textit{HST}/WFC3 brown dwarf spectra presented in Manjavacas et al. (2019)\cite{Manjavacas2019}. The self-luminous models generally show very distinct spectra from the hot Jupiter models. In particular, the self-luminous models consistently show negative values of $S_{H_{2}O}$ indicating absorption features across the full range of temperatures modeled. This is because the atmospheric thermal inversions which produce emission features can only appear in atmospheres primarily heated from above (e.g., \cite{Hubeny2003}). Additionally, at temperatures below $2000$~K where both sets of models show absorption features, the self-luminous models show consistently deeper features than the hot Jupiter models. We find this is due to the self-luminous models generally showing steeper T-P profiles, and therefore deeper absorption features, than the hot Jupiter models. We find that the hot Jupiter data are discrepant from all of the self-luminous models at $\geq10\sigma$ significance. However, at the low temperatures of observed brown dwarfs (photospheric $T<2000$~K), we find that the brown dwarf and hot Jupiter observations show similar water feature strengths. This is likely due to clouds, which can act to mute water feature strengths in both hot Jupiters and brown dwarfs at $T<2000$~K (see Supplementary Figure~\ref{fig:clouds}). This muting of water features makes the brown dwarf population appear consistent with the cloud-free hot Jupiter models between 1 and 2 microns as shown in Figure~\ref{fig:colormag}. This is because the effect of clouds in muting brown dwarf absorption features is degenerate with the steepness of their T-P profiles, as shown previously by Burningham et al. (2017)\cite{Burningham2017}. Therefore, a cloudy brown dwarf with a steep T-P profile can appear to have a similar water feature strength as a model for a cloud-free hot Jupiter with a less steep T-P profile. However, we emphasize that at temperatures $T>2000$~K, the population of observed hot Jupiters show emission features and featureless spectra ($S_{H_{2}O} \leq 0$), which are consistent with our models of hot Jupiter atmospheres but inconsistent with the absorption features shown by self-luminous objects. We also note that the scatter in the observed brown dwarf population is likely due to variation in gravity and modest variation in metallicity, as changes in the gravity can influence the cloud cover. Brown dwarfs span a much wider range of gravities than hot Jupiters but are generally expected to have much smaller differences in atmospheric composition, particularly C/O ratio\cite{Brewer2016}.

\begin{figure*}[htbp]
    \centering
    \includegraphics[width=\linewidth]{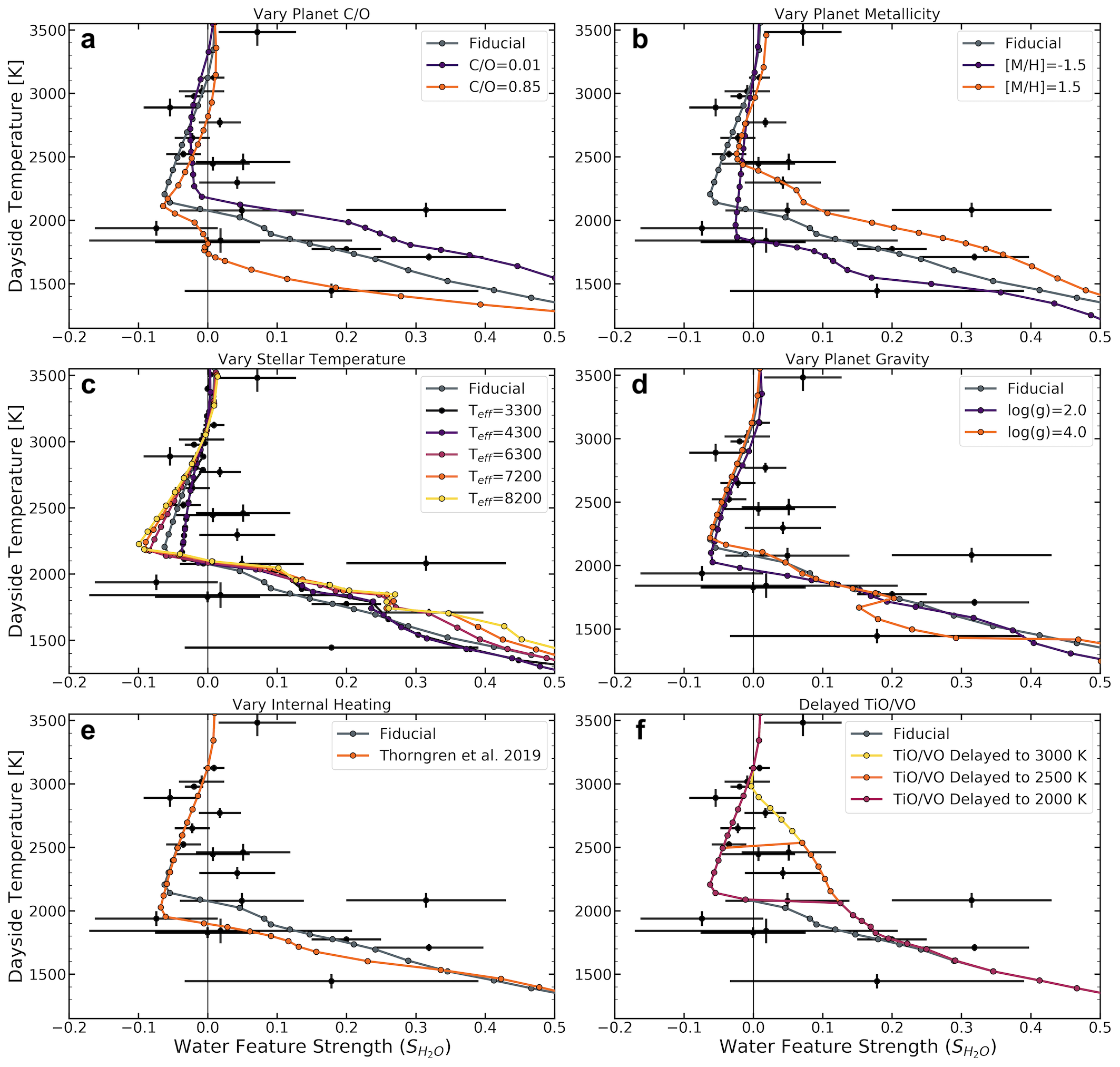}
    \caption{Diagrams illustrating the change in \textit{HST} water feature strength from models with different parameters. All diagrams show the observed hot Jupiter data as black points with $1\sigma$ error bars, while the lines show tracks for models with varying  C/O ratio \textbf{(a)}, metallicity \textbf{(b)}, stellar temperature \textbf{(c)}, gravity \textbf{(d)}, internal heating \textbf{(e)}, and the temperature to which TiO/VO opacity were ignored \textbf{(f)}. In each case all other parameters are held fixed at the fiducial model values. The error bars include uncertainties in the stellar effective temperature. We found that changing the stellar temperature, planetary gravity, and internal heating in our models had little impact on the derived water feature strengths, and changing the TiO/VO only had an impact at intermediate temperatures, but changing the atmospheric C/O ratio and metallicity can explain the diversity of observed secondary eclipse spectra.}
    \label{fig:metandco}
\end{figure*}

Although the observed population of hot Jupiter emission spectra generally matches the trends in our hot Jupiter model predictions, we find that no single model track is the best fit for all 19 of the observations. When taking each model track individually, the data are discrepant from each track at $\geq2\sigma$ significance. The fact that different data sets are best fit by models with different parameters suggests there may be one or more parameters varying between the planets. To determine which parameters can most easily explain the scatter in the observed data, we examined the water feature strength variation we could achieve through changing each of our model parameters individually. Figure~\ref{fig:metandco} shows water strengths for each individual model we examined. We found that the stellar effective temperature, planet gravity, and extent of internal heating had relatively small impacts on the predicted water feature strengths throughout the range of temperatures of the hot Jupiter population. The relatively small impact of the gravity on hot Jupiter water feature strengths is notably different than the large impact of gravity on the feature strengths of observed brown dwarfs. This difference is because the observed hot Jupiters generally show a much smaller span of gravities than the wide range represented in the population of observed brown dwarfs. Additionally, the models with TiO/VO opacity removed at different temperatures could only account for some of the scatter at intermediate temperatures and could not explain scatter at the highest or lowest temperatures, where we have observed the most precise secondary eclipse spectra. However, changing the atmospheric metallicity and C/O ratio had a significant impact on the predicted \textit{HST}/WFC3 water feature strengths. We found the observed scatter could be explained if the planets have atmospheric metallicites between $0.03-30$x solar and C/O ratios between 0.01-0.85 ($0.02-1.5$x solar). With the current data we are unable to compare each planet's specific atmospheric composition to this prediction, as even the most detailed \textit{HST} secondary eclipse observations only constrain the metallicity to within 0.5 dex and often cannot constrain the C/O ratio, or can only place an upper limit (e.g., \cite{Arcangeli2018,Mansfield2018,MikalEvans2020}). However, such variation is expected from planet formation models \cite{Mordasini2016,AliDib2017} and has been suggested by a handful of transmission spectra studies (e.g., \cite{Cridland2019,Sing2016}). The scatter we observed in emission spectra lends further support to the concept of compositional diversity among hot Jupiters.

Our hypothesis that hot Jupiters show compositional diversity can be tested through high-precision observations that cover more of the key O- and C-bearing molecules than are included in existing data sets (e.g, H$_{2}$O, CO, CO$_{2}$, and CH$_{4}$). Such observations will be possible with the upcoming \textit{James Webb Space Telescope} (\textit{JWST}) \cite{Greene2016} and stabilized, high-resolution spectrographs on large ground-based telescopes that have broad wavelength coverage \cite{Snellen2010}. Simultaneous detection of multiple molecules would lead to more precise constraints on metallicities and carbon-to-oxygen abundance ratios (and additional elemental ratios including nitrogen, etc.) \cite{Brogi2019}. Beyond testing our hypothesis, more precise compositional constraints on exoplanet atmospheres would inform our understanding of the formation and evolution processes that have produced the diverse planetary systems revealed over the last 25 years.

\end{multicols}

\thispagestyle{empty}

%\bibliography{main.bib}
%\textbf{BIBLIOGRAPHY PROBLEMS - FIGURE OUT LATER}

\begin{multicols}{2}
\section*{Methods}

\subsection*{New Observations and Data Reduction}

We reduced and analyzed \textit{HST}/WFC3+G141 spectra of six planets. At the time this study was begun, these were all of the remaining secondary eclipse data sets in the \textit{HST} archive that had not been published yet. Since we began this project, results for three planets have been published \cite{Edwards2020,Fu2020,MikalEvans2020,Pluriel2020}. In all of these cases, our reductions produced spectra consistent with the published results. Supplementary Table~3 lists the details of these observations, which included single eclipses of HAT-P-41b, KELT-7b, WASP-74b, WASP-76b, and WASP-79b, as well as five eclipses of WASP-121b.

We reduced the data using the data reduction pipeline described in Kreidberg et al. (2014)\cite{Kreidberg2014a}. We used an optimal extraction procedure \cite{Horne1986} and masked cosmic rays. To subtract the background out of each frame, we visually inspected the images to find a clear background spot on the detector and subtracted the median of this background area. The uncertainties on the measurements were determined by adding in quadrature the photon noise, read noise, and median absolute deviation of the background.

Following standard procedure for \textit{HST}/WFC3 eclipse observations, we discarded the first orbit of each visit. The spectra of each planet were binned into 14 channels at a resolution $R\approx30-40$. We also created a broadband white light curve for each planet by summing the spectra over the entire wavelength range.

We fit both the white light curves and spectroscopic light curves with a model in the form
\begin{equation}
M(t)=E(t)(cs+vt_{vis})(1-e^{-r_{1}t_{orb}-r_{2}}),
\label{eq:systematics}
\end{equation}
where M(t) is the modeled flux, E(t) is an eclipse model found using \texttt{batman} \cite{Kreidberg2015}, and the rest of the equation is a systematics model based on Berta et al. (2012)\cite{Berta2012}. In this systematics model, $c$ is a normalization constant, $s$ is a scaling factor to account for an offset in normalization between scan directions, $v$ is a visit-long linear trend, $t_{vis}$ is the time since the beginning of the visit, $r_{1}$ and $r_{2}$ are the amplitude and time constant of an orbit-long exponential ramp, respectively, and $t_{orb}$ is the time since the beginning of the orbit. For the white light curves, the free parameters in the eclipse model were the mid-eclipse time $T_{0}$ and the planet-to-star flux ratio $F_{p}$/$F_{s}$. For the spectroscopic light curves, the mid-eclipse time $T_{0}$ was fixed to the best-fit value from the white light curve and the only free parameter in the eclipse model was the planet-to-star flux ratio $F_{p}$/$F_{s}$.

The single eclipses observed for most of these planets had poor coverage of ingress and egress, so they could not constrain parameters such as the inclination to the level of precision provided by previous observations. Therefore, following best practices from previous studies (e.g.,\cite{Mansfield2018,Nikolov2018,Edwards2020,Fu2020,Pluriel2020}), all other eclipse parameters were fixed to the literature values listed in Supplementary Table~4. For the systematics model, $c$, $v$, and $s$ were allowed to vary between visits, while $r_{1}$ and $r_{2}$ were fixed to the same values for all visits. Four of the data sets (for HAT-P-41b, WASP-74b, WASP-79b, and WASP-121b) only used forward scanning instead of bi-directional scanning, so for these observations we fixed $s=1$. The first secondary eclipse observation for WASP-121b occurred two years before the other four observations and showed significant differences in the ramp shape, so we allowed this first eclipse to be fit with different values of $r_{1}$ and $r_{2}$ than the other four visits. 

The data sets for WASP-76b and WASP-79b showed additional correlated noise after applying this systematic model, so for these two data sets we tested adding an additional quadratic term to the visit-long trend. While adding this additional term was able to correct for the correlated noise, it introduced strong degeneracies between the fit parameters and the planet-to-star flux ratio. In order to avoid these degeneracies, we fit for only a linear visit-long trend in our final fit and used the divide-white method to correct for the additional noise \cite{Kreidberg2014a}.

We estimated the parameters with a Markov Chain Monte Carlo (MCMC) fit using the \texttt{emcee} package for Python \cite{Foreman2013}. The final secondary eclipse spectra for all of the planets are shown in Supplementary Figure~\ref{fig:newspectra}, and the planet-to-star flux ratio in each wavelength bin is listed in Supplementary Table~5. The white light curves had reduced chi-squared values between $1.9<\chi^{2}_{\nu}<15.2$. The spectroscopic light curves generally achieved photon-limited precision, with $\approx90$~\% of the light curves having reduced chi-squared values between $0.7<\chi^{2}_{\nu}<2.0$. However, occasional individual spectroscopic light curves had higher reduced chi-squared values between $2.0<\chi^{2}_{\nu}<3.4$. Therefore, before fitting each spectroscopic light curve we rescaled the uncertainties by a constant factor such that each light curve had $\chi^{2}_{\nu}=1$ to give more conservative error estimates. 

WASP-76 has a companion star whose spectrum is blended with that of the primary star in the WFC3 data. We corrected for the presence of this companion star using the following equation:
\begin{equation}
    F_{corr}=F_{obs}\left(1+\frac{F_{B}}{F_{A}}\right),
\end{equation}
where $F_{corr}$ is the corrected planet-to-star flux ratio in a given bandpass, $F_{obs}$ is the observed flux ratio in that bandpass including the companion star contamination, $F_{B}$ is the flux of the companion star in that bandpass, and F$_{A}$ is the flux of the primary star in that bandpass. We used ATLAS models \cite{Castelli2003} with temperatures of 6250~K and 4824~K to represent the primary star and the companion star, respectively \cite{Southworth2020}.

\subsection*{Reanalysis of Kepler-13Ab}

In addition to the six new data reductions described above, we performed a reanalysis of the emission spectrum of Kepler-13Ab. The details for the two observed secondary eclipses of Kepler-13Ab are listed in Supplementary Table~3. These data were also reduced using the data reduction pipeline described in Kreidberg et al. (2014)\cite{Kreidberg2014a}, and we again discarded the first orbit of each visit. We additionally discarded 14 of the 1008 observed spectra because they showed anomalously low fluxes in the broadband white light curve compared to the rest of the spectra. The spectrum was binned into 14 channels at a resolution of $R \approx 30-40$.

The spectrum of Kepler-13Ab was observed in stare mode. Stare mode observations commonly show one or more of three types of systematics: a visit-long trend, an L-shaped hook trend over an individual orbit, and thermal breathing \cite{Wakeford2016}. We tested including all of these components in our fit and found that only a visit-long trend was necessary to explain the systematics. Therefore, we fit both the white light curve and the spectroscopic light curves with a model in the form
\begin{equation}
    M(t)=E(t)(c+vt_{vis}).
\end{equation}
Following our method for the other data sets, the free parameters in the white light curve fit were $c$, $v$, $T_{0}$, and $F_{p}$/$F_{s}$. For the spectroscopic light curves, the free parameters were $c$, $v$, and $F_{p}$/$F_{s}$, and $T_{0}$ was fixed to the best-fit value from the white light curve. All other eclipse parameters were fixed to the literature values listed in Supplementary Table~4. The parameters $c$ and $v$ were allowed to vary between visits.

We estimated the parameters with a Markov Chain Monte Carlo (MCMC) fit using the \texttt{emcee} package for Python \cite{Foreman2013}. The final secondary eclipse spectrum for Kepler-13Ab is shown in Supplementary Figure~\ref{fig:newspectra}, and the planet-to-star flux ratio in each wavelength bin is listed in Supplementary Table~5. The white light curve had a reduced chi-squared of $\chi^{2}_{\nu}=1.33$. The spectroscopic light curves generally achieved photon-limited precision and had reduced chi-squared values between $0.94<\chi^{2}_{\nu}<1.08$.

\subsection*{Observed Dayside Temperatures and Water Feature Strengths}

As described in the main text, we determined the water feature strength ($S_{H_{2}O}$) of each observed exoplanet using Equation (1). In order to ensure that $S_{H_{2}O}$ would produce the same value for identical planets orbiting different stars, we first subtracted out the stellar signal from the observed data. We used ATLAS models\cite{Castelli2003} to calculate the stellar flux in our defined out-of-band and in-band regions. We then calculated the planetary flux in each of these regions using the equation
\begin{equation}
    F_{p}=D F_{s} \left(\frac{R_{*}}{R_{p}}\right)^{2},
\end{equation}
where $D$ is the observed secondary eclipse depth in each bandpass, $F_{s}$ is the stellar flux from the ATLAS model, and $R_{p}$ and $R_{*}$ are the planetary and stellar radius, respectively.

We measured the dayside temperature of each observed planet by fitting a blackbody to the ``out-of-band'' regions indicated in Supplementary Figure~\ref{fig:colorcalc}. Similar to previous studies\cite{Schwartz2015}, we found a linear relationship between this observed dayside temperature and the planetary irradiation temperature given by
\begin{equation}
    T_{day}=0.807_{-0.004}^{+0.008}T_{irr}+71_{-8}^{+25},
\end{equation} where $T_{irr}=T_{eff}\sqrt{R_{*}/a}$ is the irradiation temperature and $a$ is the semi-major axis.

\subsection*{Model Grid}

We created a new grid of self-consistent, 1D hot Jupiter models to compare their emission spectra to the population of observed planets. These models were generated using the Sc-CHIMERA code (validated against established brown dwarf models \cite{SaumonMarley2008} and analytic models \cite{Piskorz2018}) assuming cloud-free, radiative-convective-thermochemical equilibrium atmospheres. The models' assumption of chemical equilibrium is likely a good approximation for the highly irradiated planets that make up the majority of our observed population \cite{Kitzmann2018}. A two stream source function technique \cite{Toon1989} is employed to solve for the planetary thermal fluxes at each atmospheric level (under the hemispheric mean approximation). We modeled the stellar flux via a standard two stream approximation (for both direct and diffuse fluxes, under the quadrature approximation) assuming cosine incident angle of $0.5$,  utilizing the PHOENIX models for the stellar spectra \cite{Husser2013}. A Newton-Raphson iteration \cite{Mckay1989} is used to determine the temperature at each model layer which ensures zero net flux divergence. We include absorption cross-sections from 0.1 - 100 $\mu$m (where available) for H$_{2}$O, CO, CO$_{2}$, CH$_{4}$, NH$_{3}$, H$_{2}$S, PH$_{3}$, HCN, C$_{2}$H$_{2}$, TiO, VO, SiO, FeH, CaH, MgH, CrH, AlH, Na, K, Fe, Mg, Ca, C, Si, Ti, O, Fe$^{+}$, Mg$^{+}$, Ti$^{+}$, Ca$^{+}$, C$^{+}$, H$_{2}$, H$_{2}$-H$_{2}$/He CIA, \cite{Lupu2014, Tennyson2020, Kurucz1995, Gharib-Nezhad2019,Gharib2021}, H$^{-}$ bound-free and free-free \cite{Bell1987,John1988}, and H$_2$/He Rayleigh scattering,  and additional UV opacities for CO, SiO, and H$_2$\cite{Kurucz1995}. Pre-computed cross-sections were converted into correlated-K coefficients at a spectral resolution of 250 using a 10 point double Gauss quadrature (with half covering the top 5\% of the correlated-K cumulative distribution function) with mixed-gas  optical depths computed using the random-overlap resort-rebin framework  (e.g., \cite{Lacis1991,Amundsen2016}). Thermochemical equilibrium molecular abundances were computed using the NASA CEA Gibbs free energy minimization code \cite{Gordon1994} combined with elemental-rain out due to condensate formation (all major Si, Fe, Mg, Ca, Al, Na, and K bearing condensates are included) given the Lodders et al. (2009) \cite{Lodders2009} elemental abundances.

We parameterized the model atmospheres with a set of five parameters: the stellar effective temeprature ($T_{eff}$), the planetary gravity ($g$), the planetary metallicity ($\left[\frac{\mathrm{M}}{\mathrm{H}}\right]$), the planetary carbon-to-oxygen ratio ($\frac{\mathrm{C}}{\mathrm{O}}$), and the planetary internal temperature ($T_{int}$). Our fiducial models had the following parameter values: $T_{eff}=5300$~K, $g=10$~m/s$^{2}$, $\left[\frac{\mathrm{M}}{\mathrm{H}}\right]=0.0$, $\frac{\mathrm{C}}{\mathrm{O}}=0.55$, $T_{int}=150$~K. Models at different irradiation temperatures were created by re-scaling the incident stellar spectrum (the PHOENIX model for a given stellar effective temperature) by the ratio of the desired irradiation temperature to the bolometric temperature of a planet at 0.05 AU around a 1 solar radius star. We created hot Jupiter models with irradiation temperatures between $500-3600$~K, with step sizes of $50-200$~K.

Following Lothringer \& Barman (2019)\cite{Lothringer2019}, we calculate the absorption mean opacity $\kappa_{J}$ and the Planck mean opacity $\kappa_{B}$ at a pressure of $10^{-2}$~bar as a function of equilibrium temperature for our fiducial models. The absorption mean opacity at a given pressure $P$ is given by
\begin{equation}
    \kappa_{J}(P)=\frac{\int_{0}^{\infty}\kappa_{\lambda}(T,P)J_{\lambda}(P)\,d\lambda}{\int_{0}^{\infty}J_{\lambda}(P)\,d\lambda},
\end{equation}
where $\kappa_{\lambda}$ is the monochromatic true absorption coefficient, $J_{\lambda}$ is the mean intensity at a given wavelength, and $T$ is the local temperature in the planet's atmosphere. The Planck mean opacity is given by
\begin{equation}
    \kappa_{B}(P)=\frac{\int_{0}^{\infty}\kappa_{\lambda}(T,P)B_{\lambda}(T)\,d\lambda}{\int_{0}^{\infty}B_{\lambda}(T)\,d\lambda},
\end{equation}
where $B_{\lambda}(T)$ is the Planck function. The absorption mean opacity represents the efficiency with which the atmosphere can absorb photons, while the Planck mean opacity represents the efficiency with which the atmosphere can emit photons \cite{Lothringer2019}. Therefore, the ratio $\kappa_{J}$/$\kappa_{B}$ describes the relative efficiency of stellar absorption vs. thermal re-radiation, and a ratio $\kappa_{J}$/$\kappa_{B}>1$ generally indicates the presence of a thermal inversion in the T-P profile. The hot Jupiters in this study can generally be thought of as emitting most of their radiation at near-infrared wavelengths, whereas incoming starlight from their host stars peaks at visible wavelengths. Therefore, increasing the amount of molecules such as TiO that are optically active at visible wavelengths will increase $\kappa_{J}$, and increasing the amount of molecules such as H$_{2}$O that are optically active at near-infrared wavelengths will increase $\kappa_{B}$.

We also created subset grids as a function of irradiation temperature where a single parameter dimension was varied while all other parameters were held fixed to their fiducial model values (no cross-variance). We examined models with a stellar $T_{\mathrm{eff}}=3300$~K, 4300~K, 6300~K, 7200~K, and 8200~K; $g=1$~m/s$^{2}$ and 100~m/s$^{2}$; $\left[\frac{\mathrm{M}}{\mathrm{H}}\right]=-1.5$ and 1.5; and $\frac{\mathrm{C}}{\mathrm{O}}=0.01$ and 0.85. For models with different metallicities, elemental abundance ratios were held constant while the overall metallicity was re-scaled relative to H. We also created a model grid where the internal temperature varies with the planetary irradiation temperature following Equation~3 in ref(\cite{Thorngren2019}). Individual model tracks with irradiation temperature for each of these variations are shown in Figure~\ref{fig:metandco}.

Opacity from gaseous TiO/VO is theorized to be a driving force behind the transition between uninverted hot Jupiter atmospheres with monotonically decreasing T-P profiles and atmospheres containing thermal inversions \cite{Fortney2008}. Some previous observations of hot Jupiters have suggested that vapor TiO/VO may not be present in high-temperature atmospheres if it is condensed in cooler parts of the atmosphere (e.g., \cite{Beatty2017}). This process, known as cold-trapping, effectively works to remove TiO/VO from places in the atmosphere where vaporized TiO/VO would be expected to be present in equilibrium. In order to study the impact of potential cold-trapping, we created models where the TiO and VO opacities are artificially set to zero until a given temperature threshold. We tested models where TiO/VO opacity is zeroed out for temperatures below 2000~K, 2500~K, and 3000~K. These tracks are also shown in Figure~\ref{fig:metandco}. Supplementary Figure~\ref{fig:bestfitmods} shows the best-fit model from this complete grid for each individual data set, and Supplementary Table~6 lists the reduced chi-squared values for these best-fit models. We find that the model grid is generally able to produce good fits to the data, with the best fits to all but two data sets having reduced chi-squared values below 2.6. However, different data sets are best fit by models with different values for the atmospheric metallicity and C/O ratio, which suggests their atmospheres may have diverse compositions. 

Recent studies have suggested clouds may have an impact on the strength of molecular absorption features observed in thermal emission (e.g., \cite{Keating2019,Taylor2020}). To test the impact the presence of clouds would have on the trends in our models, we created two cloudy models. We used the cloud model of Ackerman \& Marley (2001)\cite{Ackerman2001}, as implemented by Mai \& Line (2019)\cite{Mai2019}. Both models had a constant vertical mixing strength of $10^{8}$~cm$^{2}$/s using the Zahnle et al. (2016)\cite{Zahnle2016} timescale prescription. We tested models with sedimentation efficiencies of $f_{sed}=0.1$ and $1.0$. These models are shown compared to the fiducial model in Supplementary Figure~\ref{fig:clouds}. Adding clouds acts to weaken the water feature strengths below a dayside temperature of about 2000~K, with a smaller $f_{sed}$ leading to more effective weakening. While clouds may provide a potential explanation for the weak water feature strength of HD 189733b, the lowest-temperature hot Jupiter in our population study, we find that including clouds can not generally explain the scatter we see in water feature strengths and has no impact on the feature strengths above $T_{day}=2000$~K. Our results agree with those from general circulation models, which also show that clouds have little to no impact at temperatures above $\approx2000$~K \cite{Roman2020,Parmentier2020}.

\subsection*{Self-Luminous Object Models}
\label{sec:bd}

In order to demonstrate the difference between models of hot Jupiter atmospheres (which are primarily irradiated from above by their host stars) and self-luminous atmospheres (which are primarily heated from below by the object's interior), we created a separate model grid of cloud-free self-luminous object models using the same Sc-CHIMERA code setup. We parameterized the self-luminous model atmospheres with a set of four parameters: the effective temperature ($T_{eff,bd}$), the gravity ($g$), the metallicity ($\left[\frac{\mathrm{M}}{\mathrm{H}}\right]$), and the carbon-to-oxygen ratio ($\frac{\mathrm{C}}{\mathrm{O}}$). These models thus used an identical set of parameters to the hot Jupiter models, with the exception of irradiation from within the self-luminous body instead of from an exterior star. The fiducial self-luminous models had $g=1000$~m/s$^{2}$, $\left[\frac{\mathrm{M}}{\mathrm{H}}\right]=0.0$, and $\frac{\mathrm{C}}{\mathrm{O}}=0.55$. We created models with effetive temperatures between $1000-2800$~K with a step size of $200$~K. We also created grids with metallicities of $\left[\frac{\mathrm{M}}{\mathrm{H}}\right]=-1.0$ and $1.0$. Additionally, while a gravity of $g=1000$~m/s$^{2}$ is typical for a brown dwarf, we created grids with $g=100$~m/s$^{2}$ and $g=10$~m/s$^{2}$ for direct comparison to lower-gravity hot Jupiters. We note that the self-luminous models are cloud-free and therefore likely overpredict water feature strengths at temperatures below $\approx2000$~K.

\subsection*{Data Availability Statement}

Data that support this paper's findings and its plots are available on GitHub at \href{https://github.com/meganmansfield/HSTeclipse}{this https url}. The full model grid can be found at \href{https://www.dropbox.com/sh/gfsmqlxs6l1p0st/AABXyRA9RlZawpsknXc9Ya7ra?dl=0}{this https url}.

\subsection*{Code Availability Statement}
All code used to produce findings in this paper is available on GitHub at \href{https://github.com/meganmansfield/HSTeclipse}{this https url}.

\end{multicols}

\section*{Correspondence and Requests}

Correspondence and requests for materials should be addressed to M.M.

\section*{Acknowledgements}
Based on observations made with the NASA/ESA Hubble Space Telescope, obtained from the data archive at the Space Telescope Science Institute. STScI is operated by the Association of Universities for Research in Astronomy, Inc. under NASA contract NAS 5-26555. The authors thank Elena Manjavacas, Amaury Triaud, and an additional anonymous reviewer, whose comments greatly improved the paper. M.M. acknowledges funding from a NASA FINESST grant. M.R.L acknowledges funding from NSF AST-165220, and NASA NNX17AB56G. M.R.L also acknowledges opacity information from Roxana Lupu. M.R.L., J.L.B., and J.J.F. acknowledge funding for this work from STScI grants GO-13467 and GO-14792. J.J.F. and M.R.L. acknowledge the support of NASA grant 80NSSC19K0446. J.M.D acknowledges support from the Amsterdam Academic Alliance (AAA) Program, and the European Research Council (ERC) European Union's Horizon 2020 research and innovation programme (grant agreement no. 679633; Exo-Atmos). This work is part of the research programme VIDI New Frontiers in Exoplanetary Climatology with project number 614.001.601, which is (partly) financed by the Dutch Research Council (NWO).

\section*{Author contributions statement}
M.M. reduced and analyzed the new data sets, led the data-model comparison, and wrote the manuscript. M.R.L. created the self-consistent 1D exoplanet model grids and contributed to the writing of the manuscript.
J.L.B. contributed to the conception of the population study and the writing of the manuscript.
J.J.F. contributed to the interpretation of the results and the writing of the manuscript. 
L.W. created the self-consistent 1D self-luminous model grids. 
V.P., E.M.-R.K., C.B., and J.-M.D. contributed to the interpretation of the results. E. G.-N. generated the opacities and absorption cross-sections for the 1D model grids. D.K.S. and M.L.-M. are PIs of the \textit{HST} program GO-14767 from which we obtained the new observations that were analyzed in this work. M.R.S. and G.M.R. contributed to the conception of the population study. All authors commented on the manuscript.\\
\textit{\textbf{Facilities}}: Hubble Space Telescope, Wide Field Camera 3\\
\textit{\textbf{Software}}: batman \cite{Kreidberg2015}, emcee \cite{Foreman2013}, matplotlib \cite{Hunter2007}, numpy \cite{vanderWalt2011}, pysynphot \cite{pysynphot2013}, scipy \cite{Virtanen2019}

\section*{Competing Interests}
The authors declare no competing financial interests.

\renewcommand{\refname}{Supplementary Information References}

\newpage
\section*{Supplementary Tables}
\renewcommand{\arraystretch}{1.3}
\renewcommand{\tablename}{Supplementary Table}
\setcounter{table}{0}
\renewcommand*{\thefootnote}{\fnsymbol{footnote}}

\begin{table}[htbp]
    \centering
    \begin{tabularx}{\linewidth}{>{\centering \arraybackslash}p{3.0 cm} >{\centering \arraybackslash}p{3.0 cm} >{\centering \arraybackslash}p{2.8 cm} >{\centering \arraybackslash}p{3.0 cm} >{\centering \arraybackslash}p{4.0 cm}}
    \hline
        Planet & \textit{HST} Program \# & Number of Eclipses & Observation Mode & Literature Reference \\
    \hline \hline
    CoRoT-2b & 12181\cite{DemingProp2010} & 3 & Stare Mode & Wilkins et al. (2014)\cite{Wilkins2014} \\
    HAT-P-7b & 14792\cite{BeanProp2016} & 2 & Spatial Scan & Mansfield et al. (2018)\cite{Mansfield2018} \\
    HAT-P-32Ab & 14767\cite{SingProp2016} & 1 & Spatial Scan & Nikolov et al. (2018)\cite{Nikolov2018} \\
    HD 189733b & 12881\cite{McCulloughProp2012} & 1 & Spatial Scan & Crouzet et al. (2014)\cite{Crouzet2014} \\
    HD 209458b & 13467\cite{BeanProp2013} & 5 & Spatial Scan & Line et al. (2016)\cite{Line2016} \\
    TrES-3b & 12181\cite{DemingProp2010} & 1 & Stare Mode & Ranjan et al. (2014)\cite{Ranjan2014} \\
    WASP-4b & 12181\cite{DemingProp2010} & 1 & Stare Mode & Ranjan et al. (2014)\cite{Ranjan2014} \\
    WASP-12b & 12230\cite{SwainProp2010} & 1 & Stare Mode & Stevenson et al. (2014)\cite{Stevenson2014} \\
    WASP-18b & 13467\cite{BeanProp2013} & 5 & Spatial Scan & Arcangeli et al. (2018)\cite{Arcangeli2018} \\
    WASP-33b & 12495\cite{DemingProp2011} & 2 & Spatial Scan & Haynes et al. (2015)\cite{Haynes2015} \\
    WASP-43b & 13467\cite{BeanProp2013} & 5 & Spatial Scan & Kreidberg et al. (2014)\cite{Kreidberg2014} \\
    WASP-103b & 13660\cite{ZhaoProp2014}, 14050\cite{KreidbergProp2014} & 4 & Spatial Scan & Kreidberg et al. (2018)\cite{Kreidberg2018} \\
    \hline
    \end{tabularx}
    \caption{References and \textit{HST} program numbers for the twelve planets whose spectra were taken from the literature.}
    \label{tab:published}
\end{table}

\begin{table}[htbp]
    \centering
    \begin{tabularx}{0.5\linewidth}{>{\centering \arraybackslash}p{2.5 cm} >{\centering \arraybackslash}p{2.5 cm} >{\centering \arraybackslash}p{2.5 cm}}
    \hline
        Planet & $T_{day}$ & $S_{H_{2}O}$\\
    \hline \hline
    CoRoT-2b & $1796 \pm 42$ & $0.019 \pm 0.079$ \\
    HAT-P-7b & $2772 \pm 39$ & $0.017 \pm 0.030$ \\
    HAT-P-32Ab & $1939 \pm 59$ & $-0.074 \pm 0.089$ \\
    HAT-P-41b & $2461 \pm 66$ & $0.051 \pm 0.068$ \\
    HD 189733b & $1446 \pm 57$ & $0.178 \pm 0.212$ \\
    HD 209458b & $1711 \pm 28$ & $0.319 \pm 0.079$ \\
    KELT-7b & $2447 \pm 54$ & $0.007 \pm 0.053$ \\
    Kepler-13Ab & $3484 \pm 107$ & $0.071 \pm 0.056$ \\
    TrES-3b & $1842 \pm 97$ & $0.018 \pm 0.190$ \\
    WASP-4b & $2079 \pm 62$ & $0.049 \pm 0.089$ \\
    WASP-12b & $2890 \pm 70$ & $-0.055 \pm 0.038$ \\
    WASP-18b & $2979 \pm 20$ & $-0.020 \pm 0.013$ \\
    WASP-33b & $3126 \pm 26$ & $0.009 \pm 0.015$ \\
    WASP-43b & $1775 \pm 23$ & $0.200 \pm 0.050$ \\
    WASP-74b & $2298 \pm 48$ & $0.042 \pm 0.055$ \\
    WASP-76b & $2523 \pm 27$ & $-0.035 \pm 0.025$ \\
    WASP-79b & $2083 \pm 58$ & $0.315 \pm 0.115$ \\
    WASP-103b & $3018 \pm 50$ & $-0.009 \pm 0.033$ \\
    WASP-121b & $2651 \pm 39$ & $-0.023 \pm 0.025$ \\
    \hline
    \end{tabularx}
    \caption{Computed dayside temperatures ($T_{day}$) and water feature strengths ($S_{H_{2}O}$) for each planet following Equation~\ref{eq:strength}. The errors include uncertainties in the stellar effective temperature.}
    \label{tab:colorvals}
\end{table}

\begin{table}[htbp]
    \centering
    \begin{tabularx}{\linewidth}{>{\centering \arraybackslash}p{2.0 cm} >{\centering \arraybackslash}p{2.2 cm} >{\centering \arraybackslash}p{3.0 cm} >{\centering \arraybackslash}p{3.5 cm} >{\centering \arraybackslash}p{2.7 cm} >{\centering \arraybackslash}p{1.5 cm}}
    \hline
        Planet & \textit{HST} Program \# & Date(s) of Observation & Sampling Sequence & Exposure Time [s] & Exposures per Orbit \\
    \hline \hline
        HAT-P-41b & 14767(1)\cite{SingProp2016} & 10/09/16 & SPARS10, NSAMP=12 & 81.089 & 19 \\
        KELT-7b & 14767(1)\cite{SingProp2016} & 08/18/17 & SPARS10, NSAMP=4 & 22.317 & 37 \\
        Kepler-13Ab & 13308(2)\cite{ZhaoProp2013} & 04/28/14, 10/13/14 & SPARS10, NSAMP=3 & 7.624 & 101 \\
        WASP-74b & 14767(1)\cite{SingProp2016} & 05/02/17 & SPARS25, NSAMP=4 & 69.617 & 19 \\
        WASP-76b & 14767(1)\cite{SingProp2016} & 11/03/16 & SPARS10, NSAMP=15 & 103.129 & 19 \\
        WASP-79b & 14767(1)\cite{SingProp2016} & 11/15/16 & SPARS25, NSAMP=7 & 138.381 & 13 \\
        WASP-121b & 14767(1)\cite{SingProp2016}, 15134(4)\cite{Evans2017} & 11/10/16-11/11/16, 03/12/18-03/13/18, 03/14/18, 02/03/19, 02/04/19 & SPARS10, NSAMP=15 & 103.129 & 16 \\
    \hline
    \end{tabularx}
    \caption{Observing details for the seven planets for which new data reductions were performed in this work. Numbers in parentheses next to the \textit{HST} program number indicate the number of eclipses observed in that program. Note that the spectrum of Kepler-13Ab was observed in stare mode, while all other observations were taken in spatial scanning mode.}
    \label{tab:programdeets}
\end{table}

\begin{table}[htbp]
    \centering
    \begin{tabularx}{\linewidth}{>{\centering \arraybackslash}p{3.5 cm} >{\centering \arraybackslash}p{3.5 cm} >{\centering \arraybackslash}p{2.5 cm} >{\centering \arraybackslash}p{3.0 cm} >{\centering \arraybackslash}p{3.0 cm}}
    \hline
        Planet & Period [days] &  a/r$_{*}$ & Inclination [$^{\circ}$] & r$_{\mathrm{p}}$/r$_{*}$ \\ %Mid-Transit Time [BJD] &
    \hline \hline
    HAT-P-41b & 2.694050\cite{Stassun2017} &  5.45\cite{Stassun2017} & 87.70\cite{Stassun2017} & 0.1028\cite{Johnson2017}  \\ %2454983.8617\cite{Johnson2017} &
    KELT-7b & 2.734770\cite{Stassun2017} &  5.50\cite{Stassun2017} & 83.76\cite{Stassun2017} & 0.0888\cite{Pluriel2020} \\ %2456355.2293\cite{Zhou2016} &
    Kepler-13Ab & 1.763588\cite{Beatty2017} & 4.29\cite{Beatty2017} & 86.04\cite{Beatty2017} & 0.0874\cite{Beatty2017} \\
    WASP-74b & 2.137750\cite{Stassun2017} &  4.86\cite{Stassun2017} & 79.81\cite{Stassun2017} & 0.0980\cite{Stassun2017} \\ %2456506.8926\cite{Bonomo2017} &
    WASP-76b & 1.809882\cite{Fu2020} &  4.08\cite{Fu2020} & 88.50\cite{Fu2020} & 0.1087\cite{Fu2020} \\ %2456107.8551\cite{West2016} &
    WASP-79b & 3.662380\cite{Stassun2017} &  7.03\cite{Stassun2017} & 85.40\cite{Stassun2017} & 0.1049\cite{Stassun2017} \\ %2455545.2361\cite{Bonomo2017} &
    WASP-121b & 1.274926\cite{Delrez2016} &  3.75\cite{Delrez2016} & 87.60\cite{Delrez2016} & 0.1245\cite{Delrez2016} \\ %2456635.7083\cite{Delrez2016} &
    \hline
    \end{tabularx}
    \caption{Literature values for fixed eclipse parameters in the light curve models for the seven data sets reduced in this work.}
    \label{tab:litvals}
\end{table}

\begin{table}[htbp]
    \centering
    \begin{tabularx}{\linewidth}{>{\centering \arraybackslash}p{2.5 cm} | >{\centering \arraybackslash}p{1.7 cm} >{\centering \arraybackslash}p{1.6 cm} >{\centering \arraybackslash}p{1.9 cm} >{\centering \arraybackslash}p{1.6 cm} >{\centering \arraybackslash}p{1.6 cm} >{\centering \arraybackslash}p{1.6 cm} >{\centering \arraybackslash}p{1.8 cm}}
    \hline
        Wavelength [$\mu$m] & HAT-P-41b & KELT-7b & Kepler-13Ab & WASP-74b & WASP-76b & WASP-79b & WASP-121b \\ %$F_{p}/F_{s}$ \\
    \hline \hline
    $1.120-1.159$ & $207 \pm 157$ & $284 \pm 51$ & $580 \pm 106$ & $288 \pm 67$ & $424 \pm 44$ & $12 \pm 33$ & $914 \pm 32$ \\
    $1.159-1.197$ & $461 \pm 140$ & $328 \pm 45$ & $698 \pm 103$ & $357 \pm 54$ & $589 \pm 33$ & $58 \pm 60$ & $956 \pm 32$ \\
    $1.197-1.236$ & $622 \pm 91$ & $328 \pm 49$ & $666 \pm 102$ & $304 \pm 57$ & $614 \pm 37$ & $297 \pm 54$ & $1009 \pm 33$ \\
    $1.236-1.274$ & $545 \pm 95$ & $318 \pm 54$ & $866 \pm 103$ & $310 \pm 55$ & $533 \pm 35$ & $298 \pm 64$ & $1001 \pm 29$ \\
    $1.274-1.313$ & $452 \pm 84$ & $368 \pm 48$ & $992 \pm 106$ & $429 \pm 48$ & $645 \pm 35$ & $155 \pm 49$ & $996 \pm 30$ \\
    $1.313-1.351$ & $503 \pm 79$ & $321 \pm 50$ & $932 \pm 106$ & $401 \pm 50$ & $723 \pm 33$ & $272 \pm 54$ & $1079 \pm 32$ \\
    $1.351-1.390$ & $590 \pm 81$ & $371 \pm 54$ & $821 \pm 107$ & $407 \pm 67$ & $804 \pm 33$ & $92 \pm 50$ & $1206 \pm 30$ \\
    $1.390-1.429$ & $515 \pm 82$ & $415 \pm 51$ & $879 \pm 112$ & $346 \pm 50$ & $739 \pm 36$ & $186 \pm 52$ & $1309 \pm 31$ \\
    $1.429-1.467$ & $561 \pm 84$ & $411 \pm 53$ & $857 \pm 114$ & $486 \pm 59$ & $980 \pm 37$ & $116 \pm 53$ & $1266 \pm 31$ \\
    $1.467-1.506$ & $501 \pm 87$ & $445 \pm 55$ & $895 \pm 118$ & $428 \pm 51$ & $1027 \pm 35$ & $130 \pm 58$ & $1362 \pm 32$ \\
    $1.506-1.544$ & $666 \pm 89$ & $424 \pm 56$ & $1133 \pm 124$ & $428 \pm 53$ & $993 \pm 37$ & $242 \pm 57$ & $1311 \pm 36$ \\
    $1.544-1.583$ & $613 \pm 96$ & $439 \pm 56$ & $977 \pm 128$ & $560 \pm 56$ & $1273 \pm 40$ & $185 \pm 59$ & $1370 \pm 36$ \\
    $1.583-1.621$ & $687 \pm 96$ & $447 \pm 64$ & $1363 \pm 131$ & $633 \pm 71$ & $970 \pm 45$ & $333 \pm 70$ & $1352 \pm 39$ \\
    $1.621-1.660$ & $733 \pm 106$ & $392 \pm 61$ & $967 \pm 140$ & $527 \pm 61$ & $909 \pm 47$ & $168 \pm 72$ & $1322 \pm 40$ \\
    \hline
    \end{tabularx}
    \caption{Secondary eclipse spectra for the seven planets for which new data reductions were performed in this work. All eclipse depths are in units of ppm.}
    \label{tab:spectra}
\end{table}

\begin{table}[htbp]
    \centering
    \begin{tabularx}{0.5\linewidth}{>{\centering \arraybackslash}p{2.2 cm} >{\centering \arraybackslash}p{4 cm} >{\centering \arraybackslash}p{1.5 cm} }
    \hline
    Planet & Best-Fit Model & $\chi^{2}_{\nu}$ \\
    \hline \hline
    CoRoT-2b & C/O\,$=0.85$ & 1.5 \\
    HAT-P-7b & C/O\,$=0.85$ & 0.6 \\
    HAT-P-32Ab & [M/H]\,$=-1.5$ & 1.2 \\
    HAT-P-41b & [M/H]\,$=1.5$ & 0.9 \\
    HD 189733b & C/O\,$=0.85$ & 0.5 \\
    HD 209458b & $T_{*}=6300$\,K & 1.2 \\
    KELT-7b & C/O\,$=0.85$ & 0.7 \\
    Kepler-13Ab & C/O\,$=0.01$ & 1.0 \\
    TrES-3b & Thorngren \& Fortney (2019)\cite{Thorngren2019} Internal Heating & 0.1 \\
    WASP-4b &  [M/H]\,$=-1.5$ & 0.6 \\
    WASP-12b & C/O\,$=0.01$ & 1.2 \\
    WASP-18b & C/O\,$=0.01$ & 1.8 \\
    WASP-33b & C/O\,$=0.85$ & 15.9 \\
    WASP-43b & $T_{*}=4300$\,K & 1.1 \\
    WASP-74b & [M/H]\,$=1.5$ & 0.7 \\
    WASP-76b & $T_{*}=6300$\,K & 5.9 \\
    WASP-79b & C/O\,$=0.01$ & 2.6 \\
    WASP-103b & [M/H]\,$=-1.5$ & 1.5 \\
    WASP-121b & C/O\,$=0.85$ & 1.8 \\
    \hline
    \end{tabularx}
    \caption{Best-fit models for each data set and reduced chi-squared values ($\chi^{2}_{\nu}$) for those models. In general the models produce good fits, with the best fits to all but two data sets having $\chi^{2}_{\nu} \leq 2.6$. However, different data sets are best fit by models with different values for the atmospheric metallicity and C/O ratio, which suggests their atmospheres may have diverse compositions.}
    \label{tab:bestfittab}
\end{table}

\newpage
\section*{Supplementary Figures}

\renewcommand{\figurename}{Supplementary Figure}
\setcounter{figure}{0}

\begin{figure}[htbp]
    \centering
    \includegraphics[width=\linewidth]{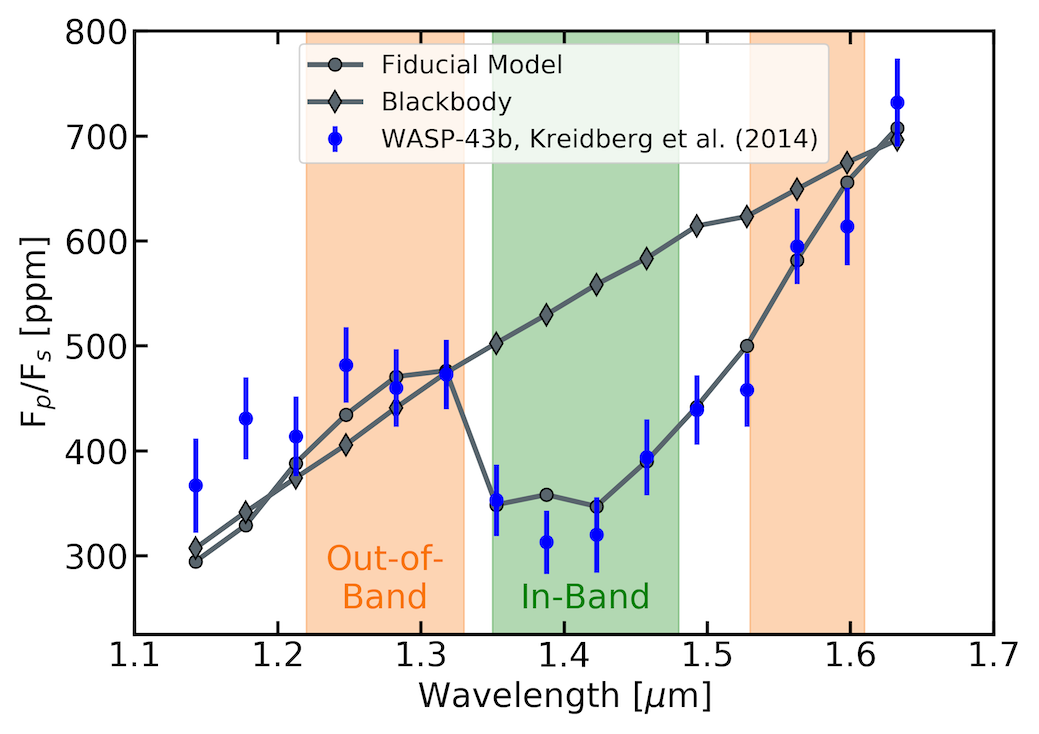}
    \caption{Construction of the \textit{HST} water feature strength metric to compare observed spectra to models. Blue points show \textit{HST}/WFC3 observations of WASP-43b \cite{Kreidberg2014}. The orange and green shaded regions indicate the spectral extent of the ``out-of-band'' and ``in-band'' flux, which are defined based on where the models in Figure~\ref{fig:models} show water features. The gray line with circular points shows the best-fit model interpolated from those in Figure~\ref{fig:models}. The gray line with diamond-shaped points shows a blackbody fit to the out-of-band flux region.}
    \label{fig:colorcalc}
\end{figure}

\begin{figure*}[htbp]
    \centering
    \includegraphics[width=\linewidth]{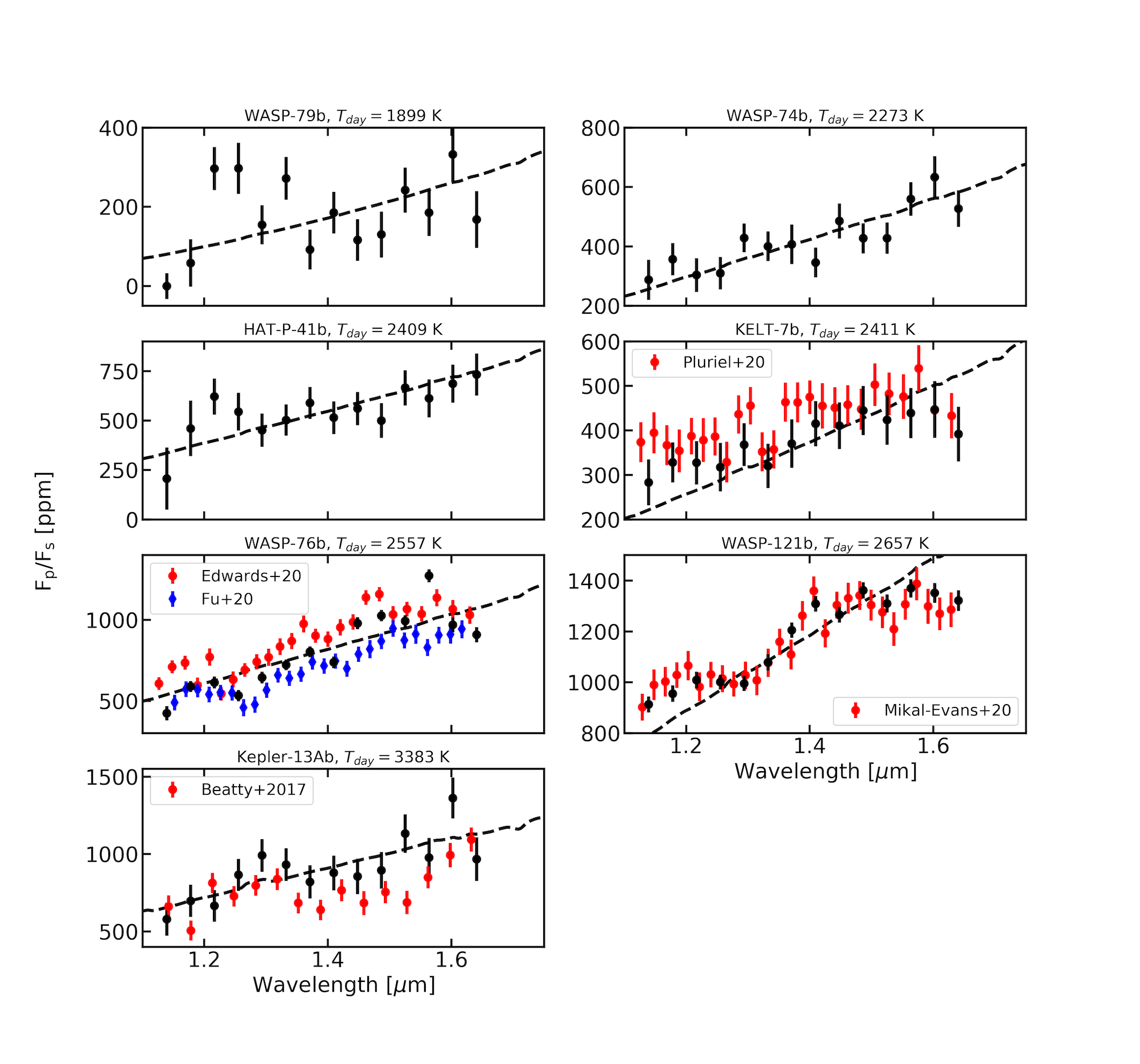}
    \caption{\textit{HST}/WFC3 secondary eclipse spectra for the six data reductions presented in this paper (black points). Black dashed lines indicate best-fit blackbody spectra, and temperatures above each plot give the corresponding dayside temperature $T_{day}$. Red and blue points show previous data reductions from the literature\cite{Edwards2020,Fu2020,MikalEvans2020,Pluriel2020}, which all show good agreement with the results presented here.}
    \label{fig:newspectra}
\end{figure*}

\begin{figure*}[htbp]
    \centering
    \includegraphics[width=\linewidth]{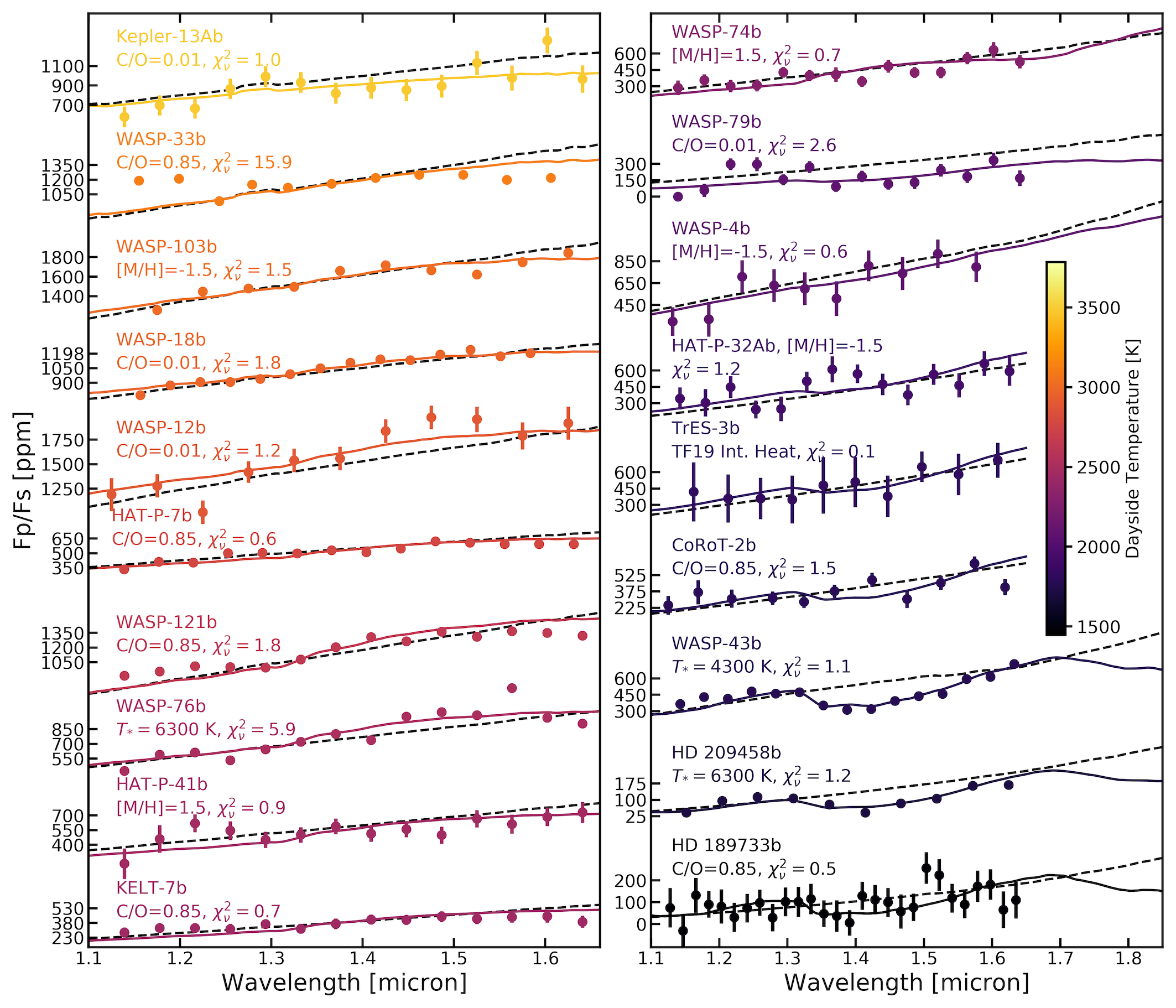}
    \caption{Same as Figure~\ref{fig:wowlookatthedata}, but showing best-fit models from the model grid instead of models from only the fiducial grid. Text below each planet name lists the model which provided the best fit for that data set and the reduced chi-squared value for that model.}
    \label{fig:bestfitmods}
\end{figure*}

\begin{figure}
    \centering
    \includegraphics[width=0.7\linewidth]{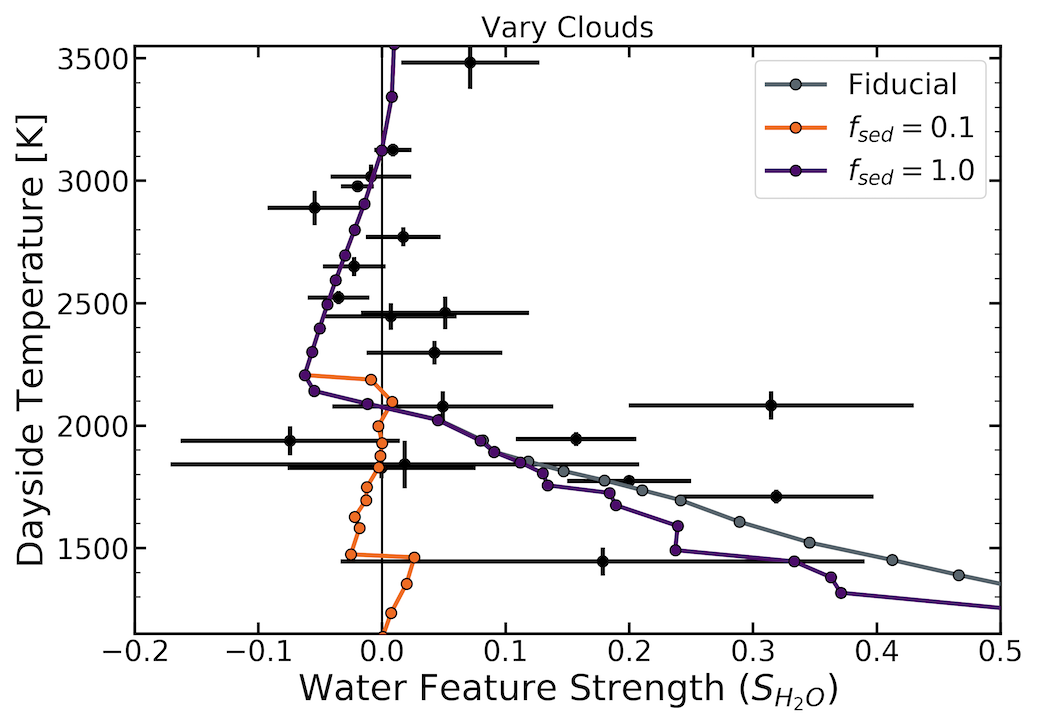}
    \caption{Change in \textit{HST} water feature strength when clouds are added to the fiducial model. The grey line shows the fiducial model, while the orange and purple lines show cloudy models with sedimentation efficiencies of $f_{sed}=0.1$ and 1.0, respectively. Adding clouds to the model effectively weakens the water feature strengths and makes the emission spectra more blackbody-like below dayside temperatures of about 2000~K. However, clouds have no effect on $S_{H_{2}O}$ at $T_{day} \geq 2000$~K because the planets' daysides are too hot for any condensation to occur.}
    \label{fig:clouds}
\end{figure}

\end{document}